\documentclass{article}
\usepackage{arxiv}
\usepackage[utf8]{inputenc} 
\usepackage[T1]{fontenc}    
\usepackage{lmodern}
\usepackage{tabularx}
\usepackage{graphicx}
\usepackage{xcolor}
\usepackage{amssymb,amsfonts,amsmath}

\usepackage{hyperref}       

\title{A high-resolution asymmetric von Hamos spectrometer for low-energy X-ray spectroscopy at the CRYRING@ESR electron cooler}

\author{
	P.~Jagodziński\thanks{\texttt{pawel.jagodzinski@ujk.edu.pl}} , D.~Bana\'{s}, M.~Pajek, A.~Kubala-Kuku\'{s}, \\ \textbf{Ł.~Jab\l{}o\'{n}ski, I.~Stabrawa, K.~Szary, D.~Sobota} \\
	Institute of Physics, Jan Kochanowski University,
	Kielce, PL-25-406, Poland\\
	\And
	A.~Warczak \\
	Institute of Physics, Jagiellonian University,
	Kraków, PL-30-348, Poland\\
	\And
	A.~Gumberidze, H.F.~Beyer, M.~Lestinsky, G.~Weber \\
	GSI Helmholtzzentrum f\"{u}r Schwerionenforschung,
	Darmstadt, D-64291, Germany\\
	\And
	Th.~St\"{o}hlker \\
	GSI Helmholtzzentrum f\"{u}r Schwerionenforschung,
	Darmstadt, D-64291, Germany\\
	GSI Helmholtz-Institut,
	Jena, D-07743, Germany\\
	\And
	M.~Trassinelli \\
	Institut des NanoSciences de Paris (INSP), CNRS,
	Paris, F-75005, France\\
}

\renewcommand{\shorttitle}{\textit{arXiv} Template}

\hypersetup{
}

\begin{document}
	\maketitle

	\begin{abstract}
		We present research program and project for high-resolution wavelength-dispersive spectrometer dedicated to low-energy X-ray spectroscopy at the electron cooler of the CRYRING@ESR storage ring, which is a~part of the international Facility for Antiproton and Ion Research (FAIR) currently being built in Darmstadt. Due to the unique shape of the electorn-ion recombination X-ray source, resulting from the overlapping of the electron and ion beams in the electron cooler, the spectrometer can work in the specific asymmetric von Hamos (AvH) geometry. In order to completely eliminate the influence of Doppler effect on the measured X-ray energies, two asymmetric von Hamos spectrometers will be installed next to the dipole magnets on both sides of the electron cooler to detect blue/red (0$^{\circ}$/180$^{\circ}$) shifted X-rays, e.g. emitted in the radiative recombination (RR) process. The X-ray-tracing Monte-Carlo simulations show that the proposed AvH spectrometer will allow to determine with sub-meV precision, the low-energy X-rays (5-10~keV) emitted from stored bare or few-electron heavy ions interacting with cooling electrons. This experimental precision will enable accurate studies of the quantum electrodynamics (QED) effects in mid-Z H- and He-like ions. 
	\end{abstract}

\keywords{Radiative recombination \and highly charged ions \and QED effects \and CRYRING@ESR electron cooler \and asymmetric von Hamos spectrometer \and high-resolution X-ray spectroscopy \and X-ray-tracing Monte-Carlo simulations}

\section{Introduction}
\label{sec:introduction}

Radiative recombination (RR) is one of the most fundamental processes occurring in collisions of highly charged ions with electrons. In this process a~free electron is captured into a~bound state of an ion with simultaneous photon emission which can be written as follows:
\begin{equation}
	A^{q+}+e(E)\rightarrow A^{(q-1)+}(n)+\hbar\omega.
\end{equation}
The emitted photon carries away the energy difference $\hbar\omega=E-E_n$ between the electron continuum state with the energy~$E$ and the electron bound state~$E_n$, with principal quantum number~$n$. The RR cross sections generally increase for lower electron energies exhibiting for bare ions $Z^2/nE$ scaling in the low-energy limit $(E<<E_n)$~\cite{pajek1992}, where for few electron ions the nuclear charge number~Z (in electron charge units) has to be replaced by an appropriate effective charge $Z_{eff}(q)$. Consequently, RR is important for low n-states of highly charged high-Z ions interacting with low-energy electrons and leads to emission of X-rays.The role of the relativistic and high multipole effects for interpretation of RR X-ray experiments at high electron energies, especially for heavy, highly charged ions has been demonstrated and discussed in~\cite{eichler2007}. 

The radiative recombination of bare or few-electron ions with cold electrons having very low energies, down to meV as it happens in the electron cooling process in ion storage rings, gives unique possibilities to study the QED effects in mid-Z and high-Z highly charged ions. In this case, the binding energies of electrons in the final bound states can be precisely measured when a~high resolution X-ray spectroscopy is applied. Consequently, the emission of X-rays from RR of highly charged ions with cold electrons in ion storage rings gives access to study the QED effects in HCI with very high precision.

Despite of the fact that the quantum electrodynamics (QED) has been tested with extremely high precision in hydrogen~\cite{niering2000, parthey2011}, the QED effects~\cite{brodsky_quantumelectrodynamics, lindren1999} for large coupling constant, up to $Z\alpha\approx0.7$ in uranium, are of fundamental interest due to the nonperturbative nature of QED corrections in this regime.  In high-Z H-like ions, the higher order QED effects, which scale with high powers of~$Z$, become more important. For the Lamb shift the dominating one-loop QED corrections, namely, the self-energy and vacuum polarization scale as $(Z\alpha)^4$, while the two-loop corrections contain terms which scale as $(\sim Z\alpha)^{5\div 7}$. For few-electron ions the QED affects additionally the electron-electron interaction~\cite{shabaev2000}. The finite nuclear size effect, increasing strongly as $(Z\alpha)^{4+2(1-(Z\alpha)^2)^{1/2}}$~\cite{brodsky_quantumelectrodynamics,shabaev2000}, limits a~precision of determination of the Lamb shift in high-Z ions due to the uncertainties in determination of nuclear charge distribution.

For this reason the studies of QED effects for mid-Z ions, in particular in the $Z$~=~20-30 range, are of special interest because here the nuclear size effect, which strongly decreases with~$Z$, becomes small, while, on the other hand, the relative precision for the determination of the Lamb shift is still acceptable for a~high energy resolution spectrometer. The precise measurements of X-ray transitions, in the context of investigations of the QED effects in mid-Z H- and He-like ions, were reported during the last decade~\cite{chantler2000, anagnostopoulos2003, bruhns2007, chantler2007, beiersdorfer2009-1, gillaspy2010, amaro2012, chantler2012, kubicek2014}. 

For bare and H-like ions with $Z$~=~20-30, the 2$P_{1/2}$~--~1$S_{1/2}$ X-ray transition are in the range of 4-10~keV. For these ions the calculated one-loop Lamb shift for H-like ions is in the range 1.6-6.4~eV, being much larger than the nuclear size effect contributing about 14-113~meV, while the two-loop QED contribution~\cite{yerokhin2003} is in the range of 1-8~meV (see Tab.~\ref{tab:contr}), respectively.

\begin{table}[!b]
	\caption{\label{tab:contr} Theoretical predictions of 2$P_{1/2}$~--~1$S_{1/2}$ energy transition, Lamb Shift of 1S$_{1/2}$ level and its contributions in H-like mid-Z ions (20~$\leqslant~\textrm{Z}~\leqslant$~30)~\cite{mohr1983, johnson1985}. All values are in eV.}
	\begin{center}
		\begin{tabular}{c|c|c|c|c|c|c|c}
			\hline
			Element & Theoretical energy &  Lamb & Self & Vac & High & Nuc & Rel \\
			& of 2$P_{1/2}$~--~1$S_{1/2}$ & Shift & En & Pol & Order & Size & Rec \\\hline
			Ca (Z~=~20) & 4101.8 & 1.634 & 1.748 & -0.129 & 0.0007 & 0.014 & 0.0005 \\
			Ti (Z~=~22) & 4968.4 & 2.259 & 2.423 & -0.188 & 0,001 & 0.024 & 0.0007 \\
			V (Z~=~23) & 5433.3 & 2.626 & 2.820 & -0.225 & 0.001 & 0.029 & 0.0007 \\
			Fe (Z~=~26) & 6956.0 & 3.973 & 4.282 & -0.365 & 0.002 & 0.053 & 0.001 \\
			Ni (Z~=~28) & 8078.2 & 5.096 & 5.507 & -0.491 & 0.003 & 0.076 & 0.002 \\
			Zn (Z~=~30) & 9287.1 & 6.431 & 6.958 & -0.646 & 0.004 & 0.113 & 0.002 \\\hline
		\end{tabular}
	\end{center}
\end{table}

\section{High resolution X-ray measurements at CRYRING@ESR storage rings}
\label{sec:cryring}

The ion storage rings equipped with an electron cooler offer unique experimental possibilities to study the RR of bare and few-electron high-Z ions with electrons. An electron cooler provides an intense beam of cold electrons which is merged with the ion beam circulating in the ring. The electrons, whose average velocity matches the average ion velocity $\langle v_e \rangle=\langle v_{ion} \rangle$ for cooling condition, are used to cool the stored ion beam by elastic electron-ion collisions, but can also be employed as an electron target to study the RR at low relative energies down to the meV range. More precisely, in the electron cooler the electron velocities in the ion frame have the so-called flattened Maxwellian distribution characterized by transverse ($T_{\perp}$)  and longitudinal ($T_{\parallel}$) electron beam temperatures with $T_{\parallel} \ll T_{\perp}$. At cooling condition the RR rates are determined by the electron transverse temperature, typically $kT_{\perp} \sim$~100~meV, which is determined by the cathode temperature in the electron gun. Generally, the RR rates for bare ions recombining with free electron having flattened velocity distribution scales as $\alpha_{RR}\sim Z^2/(kT_{\perp})^{1/2}$~\cite{pajek1992}.

\begin{figure}[!b]
	\begin{center}
		\includegraphics[width=0.65\columnwidth]{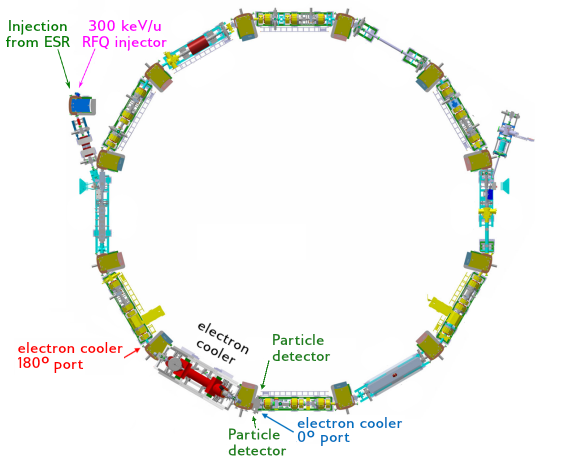}
	\end{center}
	\caption{\label{fig:cryring-view} General view of the CRYRING showing 0$^{\circ}$ and 180$^{\circ}$ ports located on the axis of the electron cooler~\cite{lestinsky2016}.}
\end{figure}
Since at the CRYRING@ESR storage ring (see Fig.~\ref{fig:cryring-view}) the adiabatic expansion of the electron beam in the cooler is applied~\cite{danared1994,danared2000}, a~reduction of the transverse electron beam temperature by two orders of magnitude, down to $kT_{\perp} \sim$~1~meV~\cite{danared2000} is reached. Taking into account that the electron density can be conserved for expanded electron beam~\cite{danared1994,danared2000}, this results in an increase of the RR rates by an order of magnitude. It is important to note that the relative electron-ion velocity in the electron cooler can be detuned from its ,,cooling'' zero-value for a~short time, which allows to measure a~dependence of both the radiative (RR) and dielectronic (DR) recombination on the relative electron energy. This is of particular interest for DR studies, which can deliver independent information on the electron beam temperatures in the electron cooler. The RR in storage rings can be studied experimentally by detecting the downcharged ions separated by the dipole magnet next to the electron cooler. In this case only the total RR rates, i.e. summed up from $n=$1 to some $n_{cut}$, which is limited by the field ionization effect in the dipole magnet, can be accessed. The counting of downcharged ions is also applied in DR studies aimed to observe narrow DR resonances, which also carry information on the electronic structure of the ion. In order to make the RR experiments state-selective, the RR photons, i.e. X-rays in case of mid- or high-Z ions, have to be observed. For bare ions stored in the ring the RR X-ray spectra measured in the electron cooler consist of a~series of RR lines from recombination of free electrons to the bound states, e.g. K-RR lines, L-RR lines, etc., and secondary X-rays from the Lyman, Balmer and higher series, due to the radiative deexcitation cascades following the initial population of higher bound states by RR process. It is important to note that the radiative cascades effectively increase the intensities of characteristic X-rays, by as much as an order of magnitude for Lyman transitions~\cite{pajek1995}.

\begin{figure}[!b]
	\begin{center}
		\includegraphics[width=0.85\columnwidth]{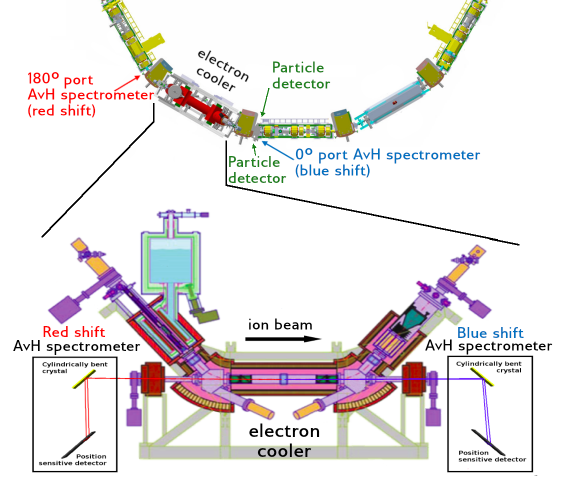}
	\end{center}
	\caption{\label{fig:cryringAvH-view} Scheme of the CRYRING@ESR showing the asymmetric von Hamos (AvH) spectrometers installed at the 0$^{\circ}$ and 180$^{\circ}$ ports located on the axis of the electron cooler~\cite{lestinsky2016}. Note that in this drawing the spectrometers are rotated by 90$^{\circ}$ (positions and dimensions of the spectrometers are not to scale).}
\end{figure}

The X-ray RR experiments at the electron cooler of the ESR storage ring at GSI have been performed with bare Au$^{79+}$ and U$^{92+}$ ions~\cite{beyer1994,liesen1994,beyer1995,gumberidze2004,gumberidze2005}, aiming at testing of QED effects. In these experiments, both the K-RR and Lyman X-rays were measured using energy-dispersive semiconductor detectors. These experiments have demonstrated the importance of the relativistic and QED effects in H-like and He-like high-Z ions. In particular, the ground-state Lamb shift was measured~\cite{gumberidze2005} with a~high precision setting one of the strongest tests of QED for high-Z H-like ions. In other experiment~\cite{banas2015} at the electron cooler a~dependence of RR rate upon the relative electron energy was measured in a~state-selective manner for U$^{92+}$  ions recombining with electrons in the relative energy range of $\pm$1~eV. This experiment showed that even for ultimately-low-energy collisions the relativistic effects are important~\cite{banas2015,reuschl2008} and, additionally, it demonstrated a~systematic enhancement of the measured RR rates for low n-states with respect to the standard RR calculations~\cite{pajek1992,bethe_quantummechanics}. This finding opens new questions for interpretation of the RR enhancement effect~\cite{horndl2005} and, on the other hand, it demonstrates that the observed RR enhancement is an important factor which can improve statistics and thus the precision of high-resolution X-ray spectroscopy applied to determine accurately the RR X-ray transition energies.

Up to now, the high-resolution wavelength-dispersive X-ray spectroscopy has been applied at the ESR only at the internal gas target. In these experiments the FOcusing Compensated Asymmetric Laue (FOCAL)~\cite{beyer2015,gassner2018} and Johann-type~\cite{trassinelli2009} X-ray crystal diffraction spectrometers have been used. At the internal gas target of the ESR, the precision of the high-resolution measurements of X-ray transition energies and, consequently of the Lamb shift, are influenced by the uncertainties in the determination of the observation angle ($\theta_{lab}$) and relativistic ion velocity $\beta=v_{ion}/c$. As we present in the following section, these drawbacks do not appear for the proposed AvH spectrometer observing the X-rays on the axis of the electron cooler simultaneously at $\theta_{lab}$~=~0$^{\circ}$ and $\theta_{lab}$~=~180$^{\circ}$, which are red- and blue- Doppler shifted photons, respectively (Fig.~\ref{fig:cryringAvH-view}).

\section{Concept of the Asymmetric von Hamos spectrometer (AvH)}
\label{sec:avhspectrometer}
As shown by X-ray-tracing Monte-Carlo simulations, which are described in more details below (see also Ref.~\cite{jagodzinski2014,jagodzinski2021}), the X-rays emitted from the electron cooler at CRYRING@ESR storage ring can be measured with a~high energy resolution, of the order of 100~meV, by using a~wavelength-dispersive X-ray spectrometer exploiting a~diffraction of photons on a~crystal to spread out the incident radiation with respect to its wavelengths on a~position sensitive X-ray detector. For low-energy X-rays (5-10~keV) the reflection-type spectrometers based on X-ray diffraction on a~flat or curved crystal geometry of Johann~\cite{johann1931}, Johansson~\cite{johansson1933} or von Hamos~\cite{vonhamos1933,vonhamos1934} can be used. The flat-crystal spectrometers generally have a~very simple construction, but they are characterized by low efficiency due to the non-focusing X-ray optics. The curved crystal geometry results in improvement of the spectrometer efficiency due to focusing of the diffracted X-rays, but can cause a~loss of energy resolution due to not well controlled deformation and waviness of the curved crystal~\cite{takagi1962}. Consequently, the von Hamos geometry, which combines a~flat crystal properties in the dispersive plane (influencing resolution) with the focusing capabilities of X-rays in the perpendicular plane (influencing intensity)~\cite{vonhamos1933,vonhamos1934}, seems to be the optimal solution to measure the X-rays from the CRYRING@ESR electron cooler.

\begin{figure}[!b]
	\begin{center}
		\includegraphics[width=0.75\columnwidth]{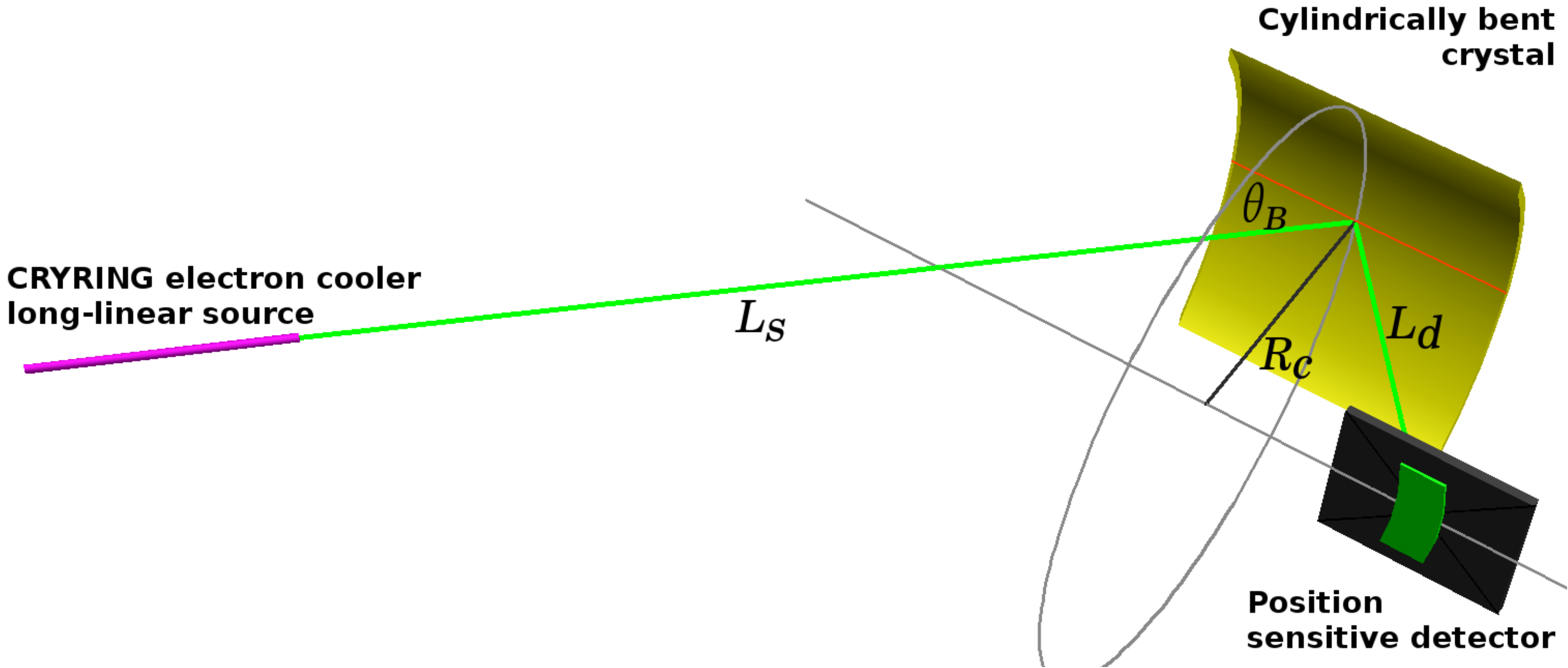}
	\end{center}
	\caption{\label{fig:sketchAvH} Sketch of the asymmetric von Hamos (AvH) spectrometer adopted for long-linear X-ray source in the electron cooler, where: $L_s$ - source-to-crystal distance, $L_d$ - crystal-to-detector distance, $R_C$ - crystal curvature, $\theta_B$ - Bragg angle.}
\end{figure}

In design of a~crystal spectrometers to be installed at the electron cooler of CRYRING@ESR the unique features of X-ray source must be taken into account. Namely, the X-ray source at electron cooler is linear, which is defined by an overlap of the ion and electron beams, with effective dimensions of about 800~mm~\cite{schuch1997}~$\times$~1~mm. For such extended X-ray source to be accepted by the X-ray diffraction spectrometer, which usually prefers a~small-size X-ray source to achieve high energy resolution, we decided to design a~spectrometer in non-standard von Hamos geometry. 

\begin{figure}[!t]
	\begin{center}
		\includegraphics[width=0.85\columnwidth]{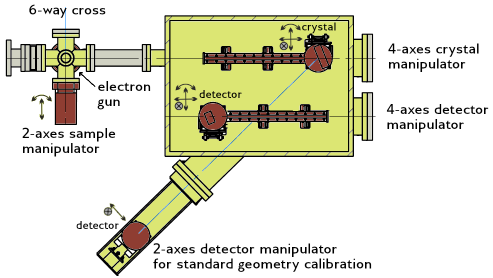}
	\end{center}
	\caption{\label{fig:koncepcjaAvH} Scheme of the asymmetric von Hamos (AvH) spectrometr (top view).}
\end{figure}

In a~standard von Hamos spectrometer~\cite{vonhamos1933,vonhamos1934} the photons emitted from a~small (close to point-like) X-ray source are diffracted at the Bragg angle~$\theta_B$ on the crystal, which is flat in the dispersive plane and cylindrically bent to the radius~$R_C$  in the focusing plane, and are measured at the focal point by a~position-sensitive X-ray detector. In the standard configuration of the von Hamos spectrometer, the source-to-crystal~$L_s$ and crystal-to-detector~$L_d$ distances are equal: $L_s=L_d=R_C/\sin\theta_B$, and both the X-ray source and the detector, are located on the same axis. In the asymmetric von Hamos geometry proposed for the X-ray measurements in CRYRING@ESR electron cooler, the RR photons are emitted from a~long ($\sim$~800~mm) linear X-ray source located about 4~m away from a~possible on-axis location of the diffraction crystal ($L_s$). That large $L_s$ distance results from a~fixed location of the 0$^{\circ}$/180$^{\circ}$ ports at the electron cooler available for installation of the spectrometers. Therefore the source-to-crystal and crystal-to-detector distances are in following relationship: $L_s\gg L_d=R_C/\sin\theta_B$ (see Fig.~\ref{fig:sketchAvH}). In this non-standard von Hamos geometry, a~high energy resolution of the spectrometer remains unchanged, mainly, due to its ,,on-axis'' location and a~small divergence ($\sim$~10~mrad) of the X-ray photons hitting the crystal.
In order to cover the 5-10~keV X-ray energy range, the asymmetric von Hamos spectrometer will be equipped with a~100~$\times$~100~mm$^2$ Si(111) crystal, cylindrically bent to a~radius of $R_C$~=~1~m. For the 3rd order of diffraction the spectrometer will cover the Bragg angle range $\theta_B$~=~36$^{\circ}$-83$^{\circ}$. The photons diffracted from a~crystal will be registered by AdvaPIX TPX3~\cite{advapix} silicon detector which consist of a~matrix of individual pixels (see Fig.~\ref{fig:koncepcjaAvH}). Both crystal and detector will be mounted on independent parallel rails.
Each rails will be equipped with 4-axes motorized manipulators, allowing for optimization of the crystal/detector angles and positions. The dispersive plane of the asymmetric von Hamos spectrometer will be oriented horizontally. The stepping precision of the linear and rotational movements will be 1~$\mu$m and 0.001$^{\circ}$, respectively, which is enough for precise adjustment of the positions and angles of the diffraction crystal and X-ray detector. The proposed AdvaPIX TPX3 detector is sensitive to the position and time of the registered X-ray photon. The detector matrix is a~square with a~side of 256 pixels of sizes 55~$\times$~55~$\mu$m$^2$. The signal processing guarantees time resolution of 1.6~ns. The timing capability of the detector with AdvaPIX TPX3 redout~\cite{advapix} will be important to eliminate the continuous X-ray background from the intense electron beam generating bremsstrahlung. This will be achieved by counting the coincidences between the X-rays and downcharged ions, separated from the main ion beam in the dipole magnet and detected by a~particle detector.

The AvH spectrometer will be able to work also in standard von Hamos geometry mode, which allow for off-line calibration of the spectrometer and checking the influence of crystal quality on the instrumental energy resolution. The off-line configuration will use a~30~keV electron gun delivering intense electron beam of a~diameter down to 150 micrometers to excite the characteristic K$\alpha$ X-rays from Fe target. The electron gun and the 2-axes target manipulator will be mounted on 6-way cross at 90 degrees angles relative to each other, whereas the detector will be installed on extended fixed-angle arm set up on the position and angle corresponding to the Fe~K$\alpha$ fluorescence line (see Fig.~\ref{fig:koncepcjaAvH}).

\begin{figure}[!t]
	\begin{center}
		\includegraphics[width=0.85\columnwidth]{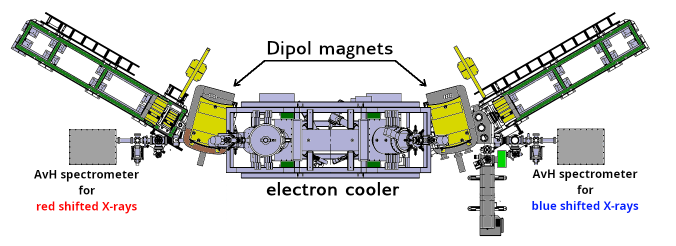}
	\end{center}
	\caption{\label{fig:ecoolerAvH-view} Scheme of the installation of the asymmetric von Hamos spectrometers at the CRYRING@ESR electron cooler.}
\end{figure}

The energy of photons observed in laboratory frame and emitted by a~moving source is modified by the Doppler effect. To eliminate the influence of the Doppler effect on measurement of the X-rays emitted from the electron cooler of CRYRING@ESR, a~pair of the asymmetric von Hamos (AvH) X-ray spectrometers will be installed before and behind the dipole magnets of the electron cooler section (Fig.~\ref{fig:ecoolerAvH-view}) in order to detect the blue- and red-shifted X-rays emitted along the electron-cooler axis. For AvH spectrometers mounted on the electron cooler axis, the measured blue- ($\theta_{lab}$~=~0$^{\circ}$) and red- ($\theta_{lab}$~=~180$^{\circ}$) Doppler shifted X-rays in the laboratory frame have the following energies:
\begin{equation}
	E_{b,r}=E_{0}[(1\pm\beta)/(1\mp\beta)]^{1/2},
\end{equation}
where $E_{0}$ is the X-ray transition energy of interest (in the moving ion frame). From this expressions it can be found that when the energies of blue- ($E_{b}$), and red- ($E_{r}$) shifted X-rays are measured, the X-ray energy in the emitter frame $E_{0}$ can be expressed as:
\begin{equation}
	E_{0}=(E_{b}E_{r})^{1/2}.
\end{equation}
This means that the X-ray transition energy in the moving ion frame can be determined from the measured energies of blue- and red-shifted photons, without the influence of the relative ion velocity~$\beta$, which usually introduce sizeable uncertainties to the Doppler effect correction.

\section{X-ray-tracing Monte-Carlo simulations}
\label{sec:simulationssoft}

\begin{figure}[!t]
	\begin{center}
		\includegraphics[width=0.85\columnwidth]{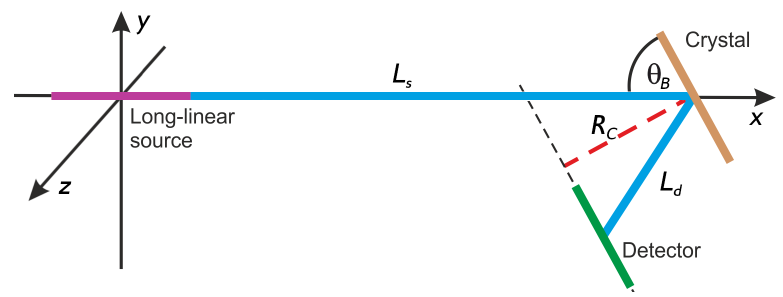}
	\end{center}
	\caption{\label{fig:schemeAvH} Geometry of the asymmetric von Hamos spectrometer shown in its dispersive plane. In this configuration, the source-to-crystal distance~$L_s$ is larger than the crystal-to-detector distance~$L_d$, which is fixed by the crystal curvature~$R_C$ and the selected Bragg angle~$\theta_B$ according to $L_d = R_C/\sin\theta_B$.}
\end{figure}

A~straightforward procedure to determine the main characteristics of the proposed AvH X-ray spectrometer, as well as to achieve better understanding of its optical properties, is the application of X-ray-tracing simulation. The Monte-Carlo method is especially well suited to study both the properties of the complex optical instruments, as well as for efficient simulations of the experiments, in particular, with moving ion beams, where due to complexity of the calculations a~fully analytical treatment is impractical. Therefore, to simulate measurements with the AvH spectrometer we used the Monte-Carlo X-ray-tracing code, which was developed to calculate the intensity distribution of diffracted X-rays on the 2D-detector for various X-ray spectrometer geometries and crystals~\cite{jagodzinski2014,jagodzinski2019,jagodzinski2021}. This X-ray-tracing code is written in C++ using Qt framework~\cite{qt}. The code tracks the trajectory of each photon emitted randomly from the X-ray source and checks whether it is diffracted by the crystal and registered by a~detector. In case when the photon reaches the crystal surface, the diffraction process described by the crystal rocking curve (from XOP2.4~\cite{delRio2011-XOP24, delRio1997, delRio1998}) is considered. The photon diffracted by the crystal is then traced on its way to the 2D-detector (see Fig.~\ref{fig:schemeAvH}).
A~shape of the X-ray source was assumed according to the conditions of the discussed experiment where the photons are emitted from the space defined by overlapping of the ion and electron beams. The simulation starts from generation of six random numbers, namely, two photon emission angles in a~spherical coordinate system, three coordinates of a~point of photon emission, and the photon energy taking into account a~natural linewidth of the transition given by a~Lorentz energy distribution. The uniform random number generators based on a~Mersenne Twister algorithm~\cite{matsumoto1998} and transformations were used to generate the random numbers having desired distributions. For X-rays emission from a~moving source, both the energy and emission angle observed in laboratory frame are subjected to the Lorentz transformation~\cite{eichler-meyerhof_relatcol}. Therefore the photon transition energy in the moving ion frame $E_{0}$ is related to photon energy in the laboratory frame $E_{lab}$ by the Doppler formula. The isotropic distribution of emitted photons in the moving ion frame is also transformed to anisotropic distribution of photons in laboratory frame. Each photon emitted from the X-ray source is traced on its path to a~crystal. A~possible diffraction point of a~photon at the crystal is determined by solving the system of equations for a~straight line photon path and a~cylindrically bent crystal surface. Additionally, the traced diffraction events have to be confined to the assumed dimensions of the crystal~\cite{jagodzinski2014}. For each calculated diffraction point on a~crystal surface, the angle~$\theta$ between photon direction and crystallographic planes is calculated. The difference $\Delta\theta=\theta-\theta_B$, where $\theta_B$ is the Bragg angle described by Bragg’s law~\cite{braggs1913}, determines the probability of X-ray diffraction according to the dynamical theory described by the diffraction profile called the crystal rocking curve~\cite{warren_xraydiffr}. In the treatment of Darwin~\cite{warren_xraydiffr, zachariasen_theoxraydiff} this approach accounts for the effects of X-ray refraction, absorption and multiple reflections in a~macroscopic perfect crystal which leads to a~corrected form of the Bragg angle~\cite{compton_xraythexp}. In this approach, the diffracted photons are scattered with a~certain probability in a~narrow angular range of $\Delta\theta$ expressed as the angular distribution of X-rays diffracted from the crystal surface. The photon diffracted from the crystal was traced on their way to the X-ray detector. The point of recording the photon on a~detector was calculated by solving the system of equations describing a~point of intersection of a~straight line diffracted photon path and a~plane representing the detector surface. In this way a~2D-distribution of photons registered by the position-sensitive detector is obtained, which is further used to obtain the energy resolution of the discussed spectrometer.

\section{Results and discussion}
\label{sec:resultsanddiscussion}

\begin{figure}[!t]
	\begin{center}
		\includegraphics[width=0.7\columnwidth]{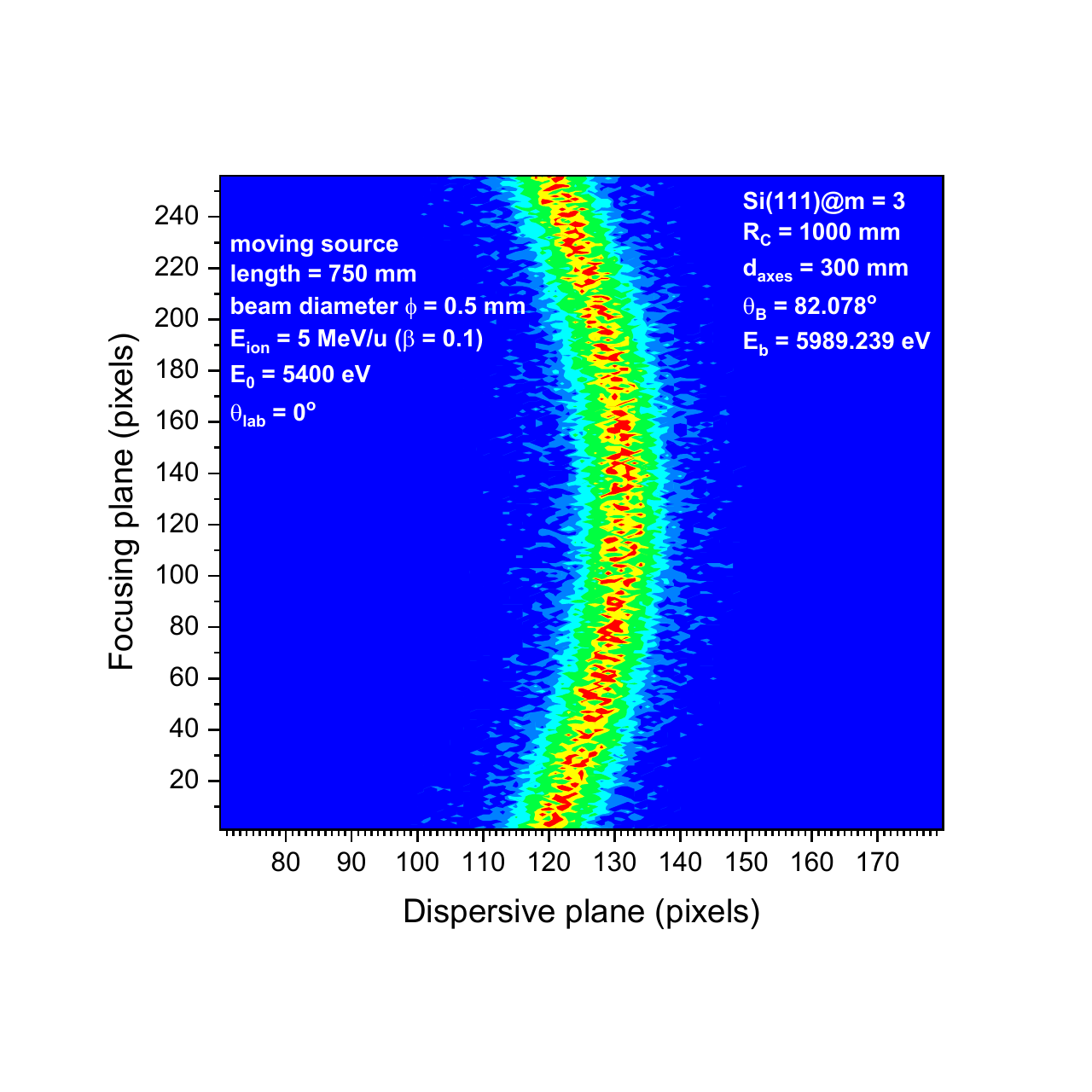}
	\end{center}\vspace{-15mm}
	\caption{\label{fig:spectrum2D} Simulated 2D-distribution of X-rays hitting the position-sensitive X-ray detector for 5.4~keV photons (in the ion frame) diffracted by the Si(111) for crystal 3rd order of diffraction for the ion beam of 5~MeV/u kinetic energy and 0.5~mm diameter.}
\end{figure}

\begin{figure}[!t]
	\begin{center}
		\includegraphics[width=0.7\columnwidth]{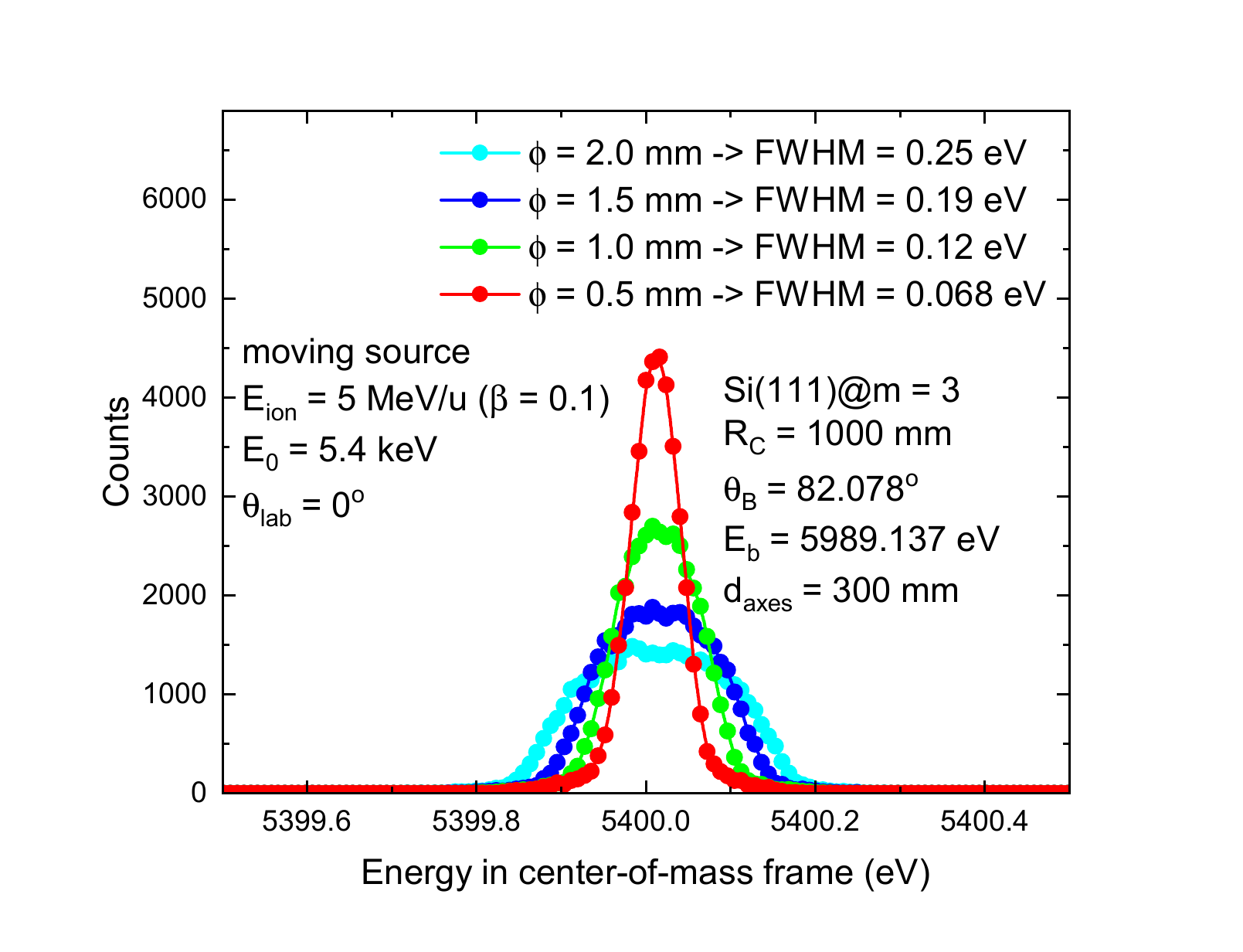}
	\end{center}\vspace{-5mm}
	\caption{\label{fig:lineProfile} Simulated line profiles of 5.4~keV photons (in the ion frame) emitted at 	forward angle ($\theta_{lab}$ = 0$^{\circ}$), diffracted by a~cylindrically bent Si(111) crystal and registered by position-sensitive X-ray detector. Fitted energy resolutions (FWHM) are shown in the figure for different ion beam diameters $\phi$.}
\end{figure}

The simulated 2D-distribution of photons emitted from moving ions, diffracted by the crystal and detected by the position sensitive detector installed in the asymmetric von Hamos spectrometer is shown in Fig.~\ref{fig:spectrum2D}. The observed pattern of detected photons is curved in the focusing plane, which is caused by the crystal bending. The energy of photons in the emitter frame is $E_{0}$~=~5400~eV. For ion kinetic energy $E_{ion}$~=~5~MeV/u corresponding to ion velocity (in units of speed of light) $\beta$~=~0.1 and for observation angle $\theta_{lab}$~=~0$^{\circ}$ (blue-shifted photons), the photon energy in the laboratory frame is $E_{b}$~=~5989.239~eV. This pattern was used to derive a~one-dimensional energy profile of the detected X-rays. This was achieved by projecting the events on the dispersion plane taking into account a~curvature of the simulated pattern, which was described by the 2nd order polynomial fitted to the simulated distribution. Such projection, which is the energy spectrum of detected X-rays, is shown in the Fig.~\ref{fig:lineProfile} for different ion beam diameters, including beam size $\phi$~=~0.5~mm corresponding to 2D-distribution in Fig.~\ref{fig:spectrum2D}. In Fig.~\ref{fig:lineProfile} simulated energy profiles for Si(111) crystal in 3rd diffraction order, bent to the radius of $R_C$~=~1000~mm and set to the Bragg angle $\theta_B$~=~82.078$^{\circ}$ for different ion beam diameters ($\phi$) in the range of 0.5-2.0~mm are presented. The FWHM widths of these profiles can be interpreted as the energy resolution of the spectrometer, because in the~simulations the natural broadening of the emission line ($\Gamma$) was not included. We found that for X-rays energy $E_{0}$~=~5.4~keV, the energy resolution varies between 68-250~meV.

\begin{figure}[!t]
	\begin{center}
		\includegraphics[width=0.7\columnwidth]{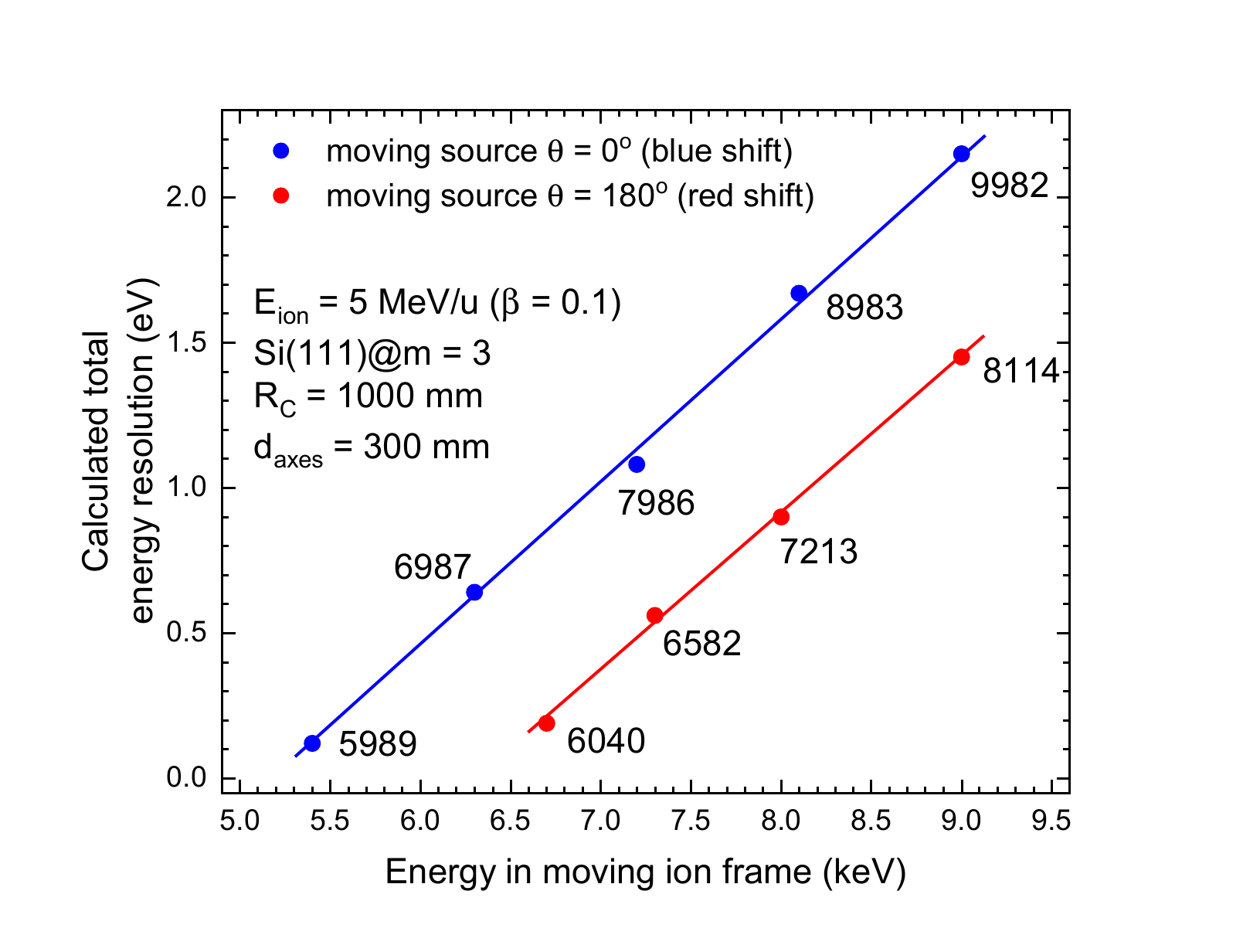}
	\end{center}\vspace{-5mm}
	\caption{\label{fig:energyres} Calculated total energy resolution of asymmetric von Hamos spectrometer for Si(111) crystals and for photon energy range of 5-9~keV (in the ion frame) for 5~MeV/u ion beam of 1~mm diameter. The values next to the points on the graph correspond to the photon energy in eV in the laboratory frame.}
\end{figure}

In order to investigate the performance of the AvH spectrometer in a~wide photon energy range of 5-10~keV, the variation of the spectrometer energy resolution was also simulated as a~function X-ray energies. The results for ion beam size of $\phi$~=~1~mm is shown in Fig.~\ref{fig:energyres}. In the range of 5-10~keV X-ray energies the energy resolution of the AvH spectrometer varies, from 0.12~eV to 2.2~eV and can be further improved down to about 70~meV by reducing ion beam diameter to 0.5~mm. However a~location of centroid of a~line profile can be determined with much better precision, in particular, when the profile is known (e.g. from simulation). By fitting such profiles, the precision of determination of a~location of its centroid, which is in fact the X-ray transition energy of interest to be measured, can be further improved down to sub-meV range for optimized ion beam size at photon energies about 5~keV. Such precision for the centroids of measured X-ray profiles was estimated by fitting the analytical model profile Asym2Sig~\cite{originlab_asym2sig} available in OriginLab software~\cite{originlab}, which reproduces the simulated line profile. Consequently, the performed simulations show that the proposed asymmetric von Hamos spectrometer can measure the energies of X-ray transitions with very high precision, down to sub-meV.

In this case, the energy of X-ray transition can be measured at 1-2~ppm precision, meaning that the QED effects can be studied at a~high precision level. It is worth noting here that the expected high precision of the order of few seb-meV, approaches the ,,natural'' linewidth of K-RR lines, which for the free-bound electron transitions to the ground state is given only by the transverse electron beam temperature being in the CRYRING@ESR electron cooler of the order of 1~meV. This shows that for the discussed case the expected precision for determination of X-ray energies nearly reaches the ultimate ,,natural'' limit set by the transverse temperature of the electron beam~\cite{pushie2014}.

\begin{figure}[!t]
	\begin{center}
		\includegraphics[width=0.7\columnwidth]{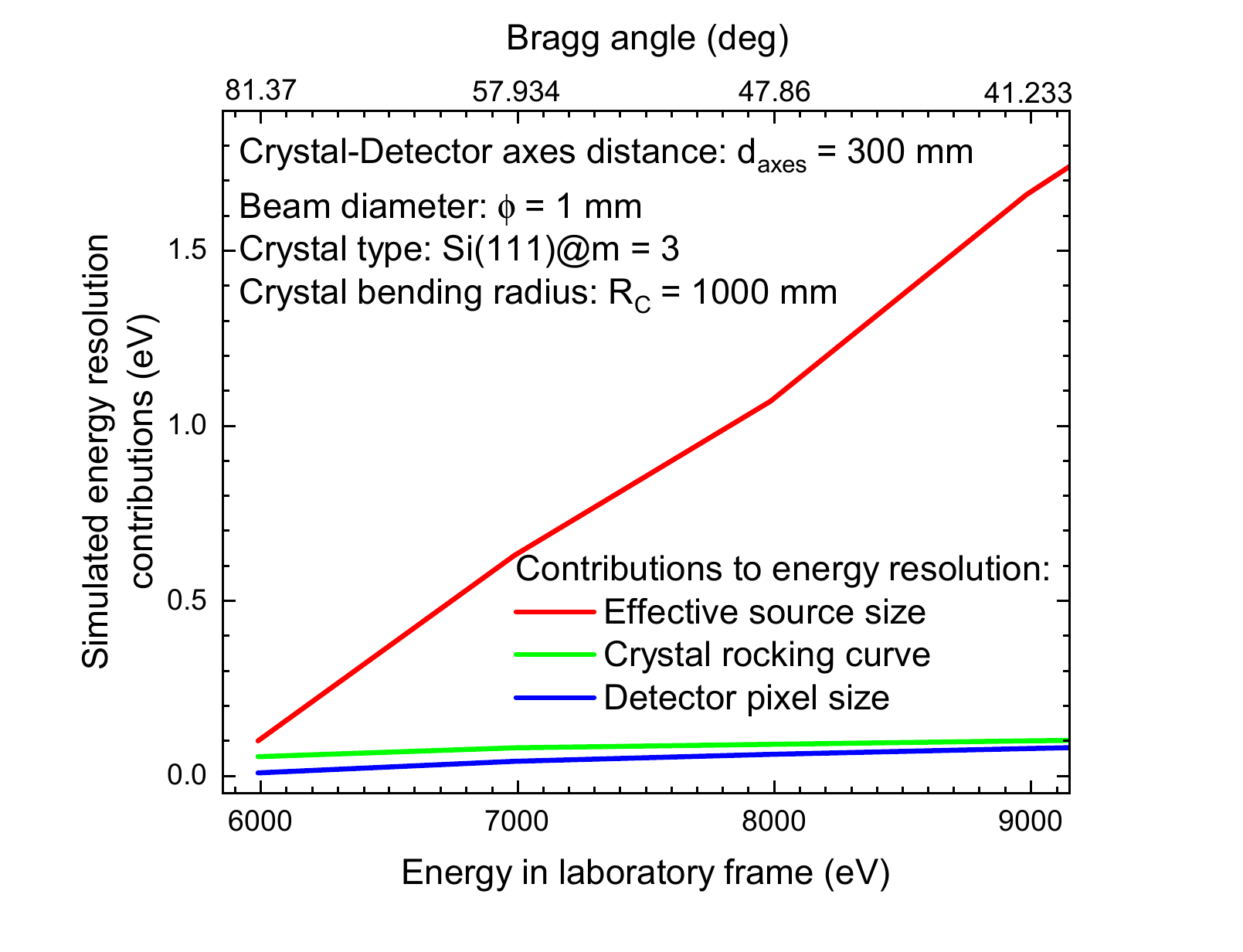}
	\end{center}\vspace{-5mm}
	\caption{\label{fig:enrescontr} Simulated contributions to the total energy resolution of AvH spectrometer: diameter size of the source (red), crystal rocking curve (green) and detector pixel size (blue).}
\end{figure}

The factors affecting the total spectrometer energy resolution are the angular divergence of the X-ray beam diffracted from the crystal, the source diameter and the spatial resolution of the position sensitive detector. Based on Monte-Carlo simulations these contributions were determined and are shown in Fig.~\ref{fig:enrescontr}. The crystal rocking curve contributes at the level of 0.055-0.095~eV, the detector pixel size contribution varies between 0.008-0.079~eV for X-ray energy range from 5~keV to 10~keV. The main contribution to the total energy resolution is caused by the transverse size of the source, which contributes at the level of 0.1-1.7~eV. The latter factor has a~decisive influence on total energy resolution of AvH spectrometer.
\begin{figure}[!b]
	\begin{center}
		\includegraphics[width=0.7\columnwidth]{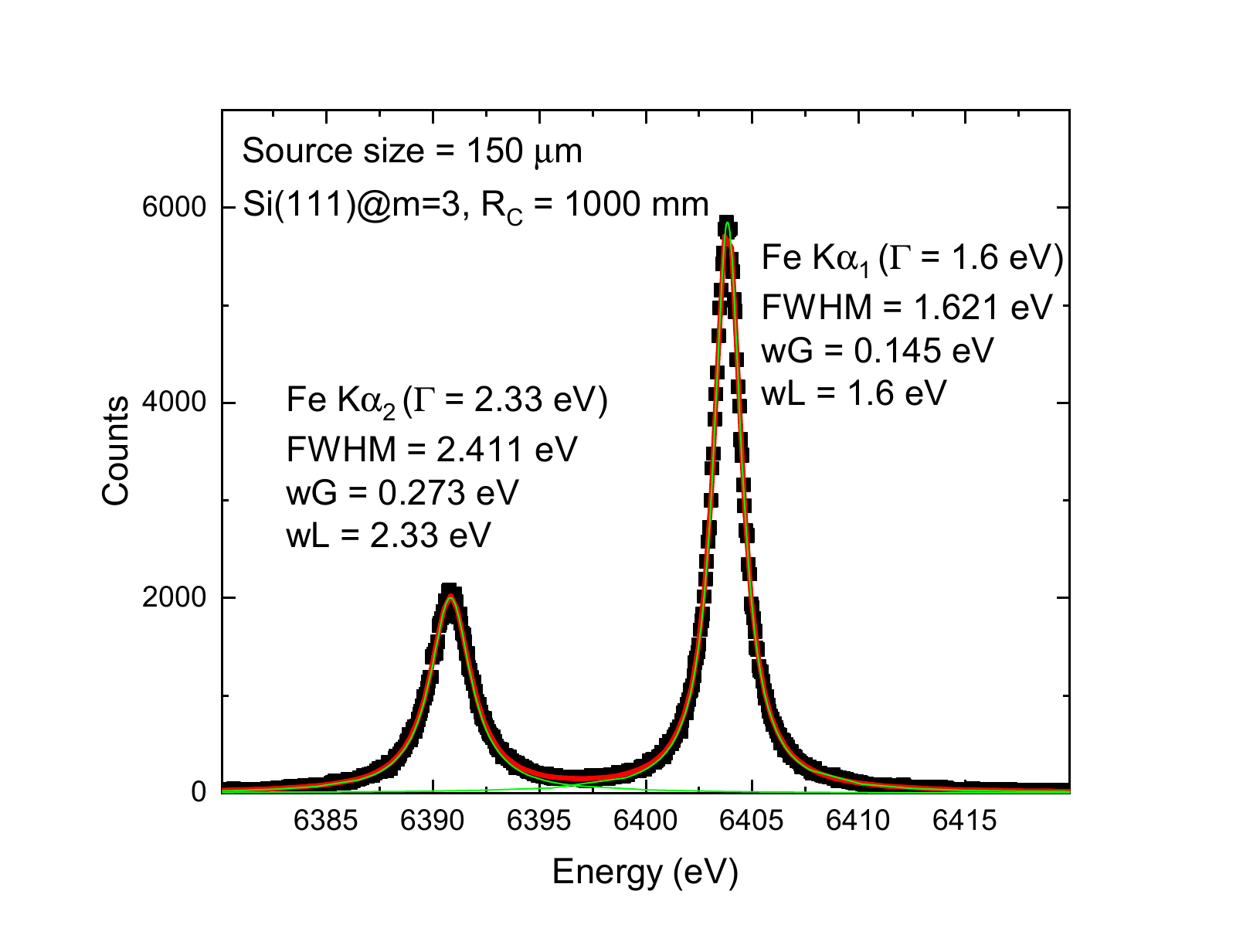}
	\end{center}\vspace{-5mm}
	\caption{\label{fig:kalibracja} Simulated Fe~$K\alpha_{1,2}$ X-ray emission lines (points) for 3rd order of diffraction with the Si(111) crystal with curvature radius of $R_C$~=~1000~mm. The solid line stand for the fit of the Fe~$K\alpha_{1,2}$ lines with the Voigt function. FWHM is the total energy resolution resulting from the convolution of the Gaussian instrumental broadening (wG) with the Lorentzian natural shape of the transition (wL).}
\end{figure}

The energy resolution of the spectrometer is also influenced by crystal waviness caused by its bending. This is a~consequence of Hooke's law according to which the curvature in one direction induces a~deformation in the perpendicular direction~\cite{krisch1991}. The proposed asymmetric von Hamos spectrometer will be equipped in fixed-angle arm (see Sec.~\ref{sec:avhspectrometer}), which allows to work in standard von Hamos geometry mode. This configuration was designed to calibrate the spectrometer as well as to check for possible crystal imperfection including its waviness, which affect the spectrometer energy resolution. Consequently, the diffraction crystal imperfection will be controlled and optimized experimentally.

The Monte-Carlo simulations were also used to determine the instrumental energy resolution of AvH, which depends only on spectrometer components as a~quality and waviness of diffraction crystal and X-ray detector, but is independent on the X-ray source properties. The instrumental energy resolution was checked for standard von Hamos geometry configuration available in the spectrometer (see Fig.~\ref{fig:koncepcjaAvH}). The simulated line profiles of X-rays of Fe~$K\alpha_{1}$ and Fe~$K\alpha_{2}$ transitions, having natural linewidth $\Gamma_1$~=~1.6~eV and $\Gamma_2$~=~2.33~eV, respectively~\cite{campbell2001}, detected by the position sensitive detector is shown in the Fig.~\ref{fig:kalibracja}. The simulated energy spectrum of iron $K\alpha$ X-ray lines exhibits clear symmetry, which is characteristic for standard von Hamos geometry. The simulated spectrum was analyzed by a~least-squares fitting procedure using Voigt function being a~convolution of the Lorentzian profiles, describing the natural X-ray transition widths, with the Gaussian function which can be interpreted as the experimental broadening of the spectrometer. For this simulation, a~size of the X-ray source, defined by the electron beam spot on a~target, generated by the electron gun, was 150~$\mu$m in diameter. In fitting, the Lorentzian widths of the $K\alpha_{1,2}$ lines were kept fixed at the values reported by Campbell and Papp~\cite{campbell2001}, while the parameters of the peak centroid and intensity, instrumental Gaussian width and linear background were used as free fitting parameters. The simulations show that the Gaussian contribution for Fe $K\alpha_{1}$ line is 0.145~eV. It should be thus noted that, the instrumental contribution to the energy resolution is greater for described standard von Hamos geometry as compared to the case of asymmetric geometry discussed in the paper. It is caused by a~bigger divergence of the photons emitted from calibrating target excited by the electron gun, which is 55~mrad as compared with 12~mrad for the RR X-ray photons emitted from the electron cooler. This calculations demonstrate that a~quality of diffraction crystal, influencing the instrumental energy resolution of the instrument can be controlled with Fe~$K\alpha$ X-rays in the AvH spectrometer. 

\begin{table}[!t]
	\caption{\label{tab:rrate} K-shell RR rate coefficients for the flattened electron beam of transverse temperature $kT_{\perp}$~=~1~meV calculated for selected bare ions with Z~=~20-30~\cite{pajek1992}.}
	\begin{center}
		\begin{tabular}{c|c|c}
			\hline
			Ion & K-shell binding energy [keV] &  K-RR rate [ph/s]\\\hline
			Ca$^{20+}$ & 5.470 & 8.1$\times$10$^5$\\
			Ti$^{22+}$ & 6.626 & 9.8$\times$10$^5$\\
			Fe$^{26+}$ & 9.278 & 1.4$\times$10$^6$\\
			Ni$^{28+}$ & 10.775 & 1.5$\times$10$^6$\\
			Zn$^{30+}$ & 12.389 & 1.8$\times$10$^6$\\\hline
		\end{tabular}
	\end{center}
\end{table}

In order to estimate the expected count rate of the K-shell RR photons in the planned experiment using the designed asymmetric von Hamos spectrometer, the needed K-shell RR rate coefficients were calculated following Ref.~\cite{pajek1992}. For example, for 10$^8$ of mid-Z bare ions stored in the ring and an electron beam density in the electron cooler of 10$^7$ cm$^{-3}$~\cite{danared2000}, the expected emission rate of K-shell RR X-rays for bare mid-Z ions, e.g. Ca$^{20+}$ ions, is of the order of 10$^6$ photons per second (see Tab.~\ref{tab:rrate}). The estimated efficiency of the asymmetric von Hamos spectrometer calculated on the basis of Monte-Carlo X-ray-tracing simulations as the ratio of the photons number emitted by the source in the direction the crystal to the number of photons registered by the detector is at the level of 1~$\times$~10$^{-7}$. Therefore, the K-RR X-ray detection count rate is expected to be of the order of 100~photons/hour. This shows that the AvH spectrometer will allow to measure the QED effects for mid-Z ions with sub-meV accuracy, being thus sensitive to two-loop QED effects.

\section{Summary}
In this paper, the research program and design of a~high energy resolution asymmetric von Hamos spectrometer for low energy (5-10~keV) X-ray spectroscopy experiments at CRYRING@ESR electron cooler is presented. Using Monte-Carlo X-ray-tracing simulations the main characteristics of the spectrometer were obtained. Performed simulations show that the QED effects in mid-Z ($Z=$~20-30) highly charged ions can be studied with an accuracy of sub-meV for photon about 5~keV enabling sensitivity to two-loop QED contributions.

\section*{Acknowledgements}
This research was funded in whole or in part by National Science Centre, Poland, under Grant No. DEC-2022/06/X/ST2/00453. For the purpose of Open Access, the author has applied a~CC-BY public copyright licence to any Author Accepted Manuscript (AAM) version arising from this submission.

\bibliographystyle{unsrt}
\bibliography{pj-bibliography.bib}

\end{document}